\documentclass[journal=aamick,manuscript=article]{achemso}

\usepackage[version=3]{mhchem}
\usepackage{graphicx}
\usepackage{pdflscape}
\usepackage{amsmath}

\usepackage{multirow}
\usepackage{multicol}
\usepackage{array}
\usepackage{booktabs} 
\usepackage{longtable}
\usepackage{xspace}
\usepackage{xcolor}

\usepackage{hyperref}
\hypersetup{colorlinks=true,linkcolor=blue,citecolor=blue,urlcolor=blue}
\usepackage{xcolor}
\usepackage{bm}
\usepackage{subcaption}
\usepackage{nicefrac}
\usepackage[separate-uncertainty=true]{siunitx}
\DeclareSIUnit\angstrom{\textup{~\AA}}

\newcommand{\ie}{\textit{i.\,e.}~}

\newcommand{\pfunit}[1]{#1Wm$^{-1}$K$^{-2}$}
\newcommand{\klunit}[1]{#1Wm$^{-1}$K$^{-1}$}
\newcommand{\sgunit}{$\Omega^{-1}$m$^{-1}$}

\author{M. Vallinayagam}
\affiliation{IEP,TU Bergakademie Freiberg, 09599 Freiberg, Germany,\\ Faculty of Physics, University of Applied Sciences, Friedrich-List-Platz 1, D-01069 Dresden, Germany}

\author{A. E. Sudheer}
\affiliation{Department of Sciences, Indian Institute of Information Technology Design and Manufacturing Kurnool Andhra Pradesh, 518007}

\author{A. Kumar}
\affiliation{Theory and Simulations Laboratory, Theoretical and Computational Physics Section, Raja Ramanna Centre for Advanced Technology, Indore 452013, India\\Homi Bhabha National Institute, Training School Complex, Mumbai 400094, India}

\author{G. Tejaswini}
\affiliation{Department of Sciences, Indian Institute of Information Technology Design and Manufacturing Kurnool Andhra Pradesh, 518007}

\author{M. Posselt}
\affiliation{Helmholtz-Zentrum Dresden-Rossendorf, Institute of Ion Beam Physics and Materials Research, Bautzner Landstraße 400, 01328 Dresden, Germany}

\author{C. Kamal}
\affiliation{Theory and Simulations Laboratory, Theoretical and Computational Physics Section, Raja Ramanna Centre for Advanced Technology, Indore 452013, India\\Homi Bhabha National Institute, Training School Complex, Mumbai 400094, India}

\author{D. Murali}
\affiliation{Department of Sciences, Indian Institute of Information Technology Design and Manufacturing Kurnool Andhra Pradesh, 518007}
\email{dmurali@iiitk.ac.in}

\author{M. Zschornak}
\affiliation{IEP,TU Bergakademie Freiberg, 09599 Freiberg, Germany,\\ Faculty of Physics, University of Applied Sciences, Friedrich-List-Platz 1, D-01069 Dresden, Germany}
\email{matthias.zschornak@htw-dresden.de}

\title[SbXY Janus layer] {Thermoelectric properties of SbXY (X = Se, Te; Y = Br, I) Janus layers}
\abbreviations{}

\begin{document}


\begin{abstract}
We report a comprehensive investigation of the thermoelectric properties of SbXY (X = Se, Te; Y = Br, I) Janus layers (JL) using spin-polarized first-principles calculations.
Ab initio molecular dynamics confirm that the 1T phase (\texttt{Pm31}) remains stable up to 1000 K, excluding any phase transitions. The calculated mean-square displacement further evidences the structural robustness. The thermal conductivity is strongly suppressed in Br-containing layers due to enhanced Fröhlich interactions between optical and acoustic phonons. Electronic structure calculations reveal indirect band gaps of 1.1–1.3 eV, with valence and conduction bands dominated by the $p$-orbitals of halogen/chalcogen and of Sb, respectively. The carrier effective mass highlights anisotropic transport with lighter electrons being more mobile, while holes dominate the power factor, which attains values on the order of $10^{-3}$ \pfunit{}. Direction-dependent transport indicates superior thermoelectric performance along the $xx$ direction, with negligible contribution along $yy$. The Figure of Merit reaches 0.6 at 1000 K in hole-doped SbSeBr, demonstrating strong potential for high-temperature applications. Our results reveal that the SbXY JLs, particularly SbSeBr, emerge as promising candidates for next-generation thermoelectric devices at elevated temperatures.
\end{abstract}

\section{\label{sec:intro}Introduction}
Growing needs for fossil fuel replacement to tackle climate change and meet the energy crisis require sound technologies for continuous generation and supply of energy by utilizing renewable sources. The direct conversion of heat into electricity is one of the promising solutions. However, designing and synthesizing materials that potentially convert temperature gradients into electricity is a challenging problem~\cite{Yan2022, Powell2025}. The performance of heat-to-electricity converting thermoelectric (TE) materials is determined by a dimensionless parameter, called the Figure of Merit $ZT$, defined by relating the materials' characteristics, Seebeck coefficient $S$, electrical conductivity $\sigma$, thermal conductivity $\kappa$, and the operating temperature as \cite{snyder08_natmat, Gorai2017, Shu2024, Xiao2018_natmat, Cao2023_eSci, binbin2021_sci, Mamani2024}: 
\begin{equation}
    ZT~=~S^2 \sigma T/\kappa \label{eq:ZT}
\end{equation}
where the thermal conductivity $\kappa$ roots from both electron ($\kappa_e$) and phonon ($\kappa_l$) contributions, \ie $\kappa~=~\kappa_e+\kappa_l$. A good thermoelectric material should possess as high $S$ and $\sigma$ as possible while maintaining low $\kappa$. The key factor to tune the parameters $S$, $\sigma$, and $\kappa$ is the concentration of charge carriers $n$. $n$ is related to the Seebeck coefficient by $S \propto n^{-2/3}$, to the electrical conductivity by $\sigma \propto n$, and to the electronic thermal conductivity by $\kappa_e \propto n$~\cite{snyder08_natmat}. Consequently, increasing $n$ enhances $\sigma$ and partially $\kappa$, but diminishes $S$. On the other hand, the increasing $\sigma$ further boosts up $\kappa$ by Wiedemann–Franz law~\cite{snyder08_natmat, DONG2023}. Hence, optimizing $ZT$ is a balancing act between maximizing both $S$ and $\sigma$ while minimizing $\kappa$~\cite{Bai2023}. 

Dimensionality reduction has emerged as a viable strategy to optimize $ZT$, owing to the increased electronic density of states near the Fermi level. For example, one-dimensional Bi$_2$Te$_3$ wires are theoretically shown to achieve a $ZT$ as high as 14~\cite{Hicks1993}. The reduction of three-dimensional (3D) lattices to two-dimensional (2D) systems has been shown to improve $S$~\cite{Parker2013}. Beyond quantum confinement, thermoelectric materials composed of heavy elements such as Bi~\cite{Gzhang2015, Shuai2016, C6TA04240F, Ningwang2021, Sujata_PCCP_2024}, Sb~\cite{Poonam2023, poonam2024jmca, Bafekry2021, SDGuo2017, Bafekry2020, Chu2023, SubarnaDas_2025}, and Pb~\cite{Mamani2024, YGan2021, PJin2022} are particularly attractive due to enhanced phonon scattering that further reduces $\kappa$. Also, chalcogen-containing 2D layers such as PbTe, SnTe, and GeTe exhibit low $\kappa_l$ due to acoustic-optical phonon modes interaction~\cite{Snair2023, JPark2024}.

In recent years, Sb-based 2D layers have attracted significant attention due to their tunable properties, enabling a wide range of applications such as photocatalytic energy production~\cite{ASE2024PCCP}, spintronics via Rashba spin splitting~\cite{YagmurcukardesAPR, Hu2025, Wu2025nano, Nwao2025}, photovoltaics~\cite{MAVLONOV2020}, optoelectronics~\cite{Liteng2025}, and photonics~\cite{Matthew2021}. Novel 2D Sb-based compounds, often designed in combination with other $p$-block elements, are actively investigated for thermoelectric applications by theoretical~\cite{poonam2024jmca, Bafekry2021, Bafekry2020, SDGuo2017, Chu2023, Kaur_2022} and experimental~\cite{Kimberly24} studies. Particularly, the Sb-based Janus layer (JL) sandwiches the Sb atomic layer between two different anionic atomic layers, forming either 2H or 1T phases, breaking the inversion symmetry and crystallizing in space group \texttt{P3m1}~\cite{Chu2023, ASE2024PCCP}. The SbTeI JL is shown to be structurally stable by density functional theoretical simulations in both 2H and 1T symmetries~\cite{Chu2023} at 0 K. Additionally, our recent theoretical results depict the structural stability of SbSeI, SbSeBr, and SbTeI JLs in the 1T phase, so far at 0 K~\cite{ASE2024PCCP}. 

The exploration of Sb JLs, particularly composed of chalcogenides and halides, has recently emerged as a rapidly growing field in 2D materials research. However, a limited number of chalcogenide-halide JLs, only Bi-based, have been experimentally realized to date. For example, BiTeI nanosheets with an average thickness of 10-20 nm were first successfully synthesized via electrochemical exfoliation \cite{Antonatos2021}. To overcome the challenges associated with mechanical and electrochemical exfoliation of BiTeI, liquid-phase exfoliation has been proposed and utilized to produce BiTeI flakes with an average thickness of 3 nm from bulk crystals \cite{Bianca22}. Moreover, a few triple layers of BiTeCl and BiTeBr with thicknesses below 10 nm have been synthesized by converting van der Waals epitaxy of Bi$_2$Te$_3$ sheets on sapphire at elevated temperatures in the presence of chemically reactive BiCl$_3$ or BiBr$_3$ vapors \cite{Hajra2020}. Recently, electrochemical exfoliation has also been extended to other BiTeX (X = Cl, Br, I) systems \cite{Wu2025nanoscale}. 

Although epitaxial growth has been successful for Bi-based JLs, similar methods have been reported to fail for Sb-based JLs \cite{Hajra2020}. To the best of our knowledge, the structural stability of the 2H and 1T phases of SbXY (X = Se, Te; Y = Br, I) JLs at elevated temperatures remains an open question, as the experimental synthesis of these layers is still under development and has yet to be realized. While the structural, optical, and catalytic properties of these systems have been extensively examined by theoretical modeling, the charge carrier mobility has not yet been addressed in detail, necessitating careful investigation of the effective mass of charge carriers. The effect of cation substitution on the thermoelectric performance of MTeI (M = Bi, Sb) has so far been investigated only under the assumption of various constant relaxation times, without considering carrier mobility explicitly \cite{SDGuo2017}. Furthermore, the reported relaxation times in SbSeBr and SbSeI JLs, ranging from \SIrange{0.0005}{0.003}{\pico\second}, appear to be implausibly small \cite{poonam2024jmca}.

Motivated by above mentioned open problems about SbXY (X = Se, Te; Y = Br, I) JLs, this study presents a comprehensive investigation of their TE properties. Specifically, we evaluate i) the effective mass, mobility, and relaxation time of charge carriers within the deformation potential framework, ii) the lattice thermal conductivity using anharmonic lattice dynamics, and iii) the transport properties through linearized Boltzmann transport equations under the constant relaxation time approximation. The TE performance of SbXY JLs is then assessed over a wide range of temperatures and carrier concentrations. In addition, the structural stability of these JLs is examined via self-consistent phonon calculations up to 1000 K. Outlining the computational details in Sec. \ref{sec:method} and elaborating the electronic properties in Sec. \ref{sec:elecpro}, our results are discussed for mobility and relaxation time of carriers (Sec. \ref{sec:mutau}), structural stability (Sec. \ref{sec:phonon}), thermal conductivity (Sec. \ref{sec:tc}) and TE properties (Sec. \ref{sec:TEprop}, and \ref{sec:TEeffi}).

\section{\label{sec:method}Methodology}
Density functional theory (DFT) based simulations to relax the structures and ground state energy prediction are carried out using the Vienna ab-initio package~\cite{kresse1996, Kresse1996_prb}. The Perdew-Burke-Ernzerhof (PBE) functional within the generalized gradient approximation (GGA)~\cite{pbe} is applied to treat the exchange-correlation interactions of electrons. Plane-waves with cutoff energy of \SI{600}{\electronvolt} are used to foster the electron-ion interactions within the projector-augmented wave (PAW) method~\cite{Blochl1994, Kresse1999}. The ground state energy of the layers is determined with \SI{e-6}{\electronvolt} accuracy when the forces on atoms are less or equal to \SI{e-3}{\electronvolt\per\angstrom}. All integrations in the reciprocal space are carried out on Monkhorst-Pack \textbf{k}-point grid of size $16\times16\times1$~\cite{monkhorst1976}. The lattice parameters are adopted from the previous report in Ref.~\cite{ASE2024PCCP}. The Boltzmann transport equation (BTE) is solved using the BoltzTraP code~\cite{btp2} to investigate the thermoelectric properties of SbXY JLs. The DFT band energies are interpolated using a Fourier-based method along with their derivatives, see Eq.(1) in Ref.~\cite{btp2}. Then the quasi-particle energies are analyzed to solve the BTE under the constant relaxation time approximation (CRTA). The formulation for electrical conductivity $\sigma$, Seebeck coefficient $S$, and charge-carrier thermal conductivity $\kappa_e$ is provided in Sec. 1 in SM. Further details of formulation and its implementations can be obtained from Ref.~\cite{btp2, btp}.

\begin{figure*}[tb]
 \centering
 \includegraphics[width=1\textwidth]{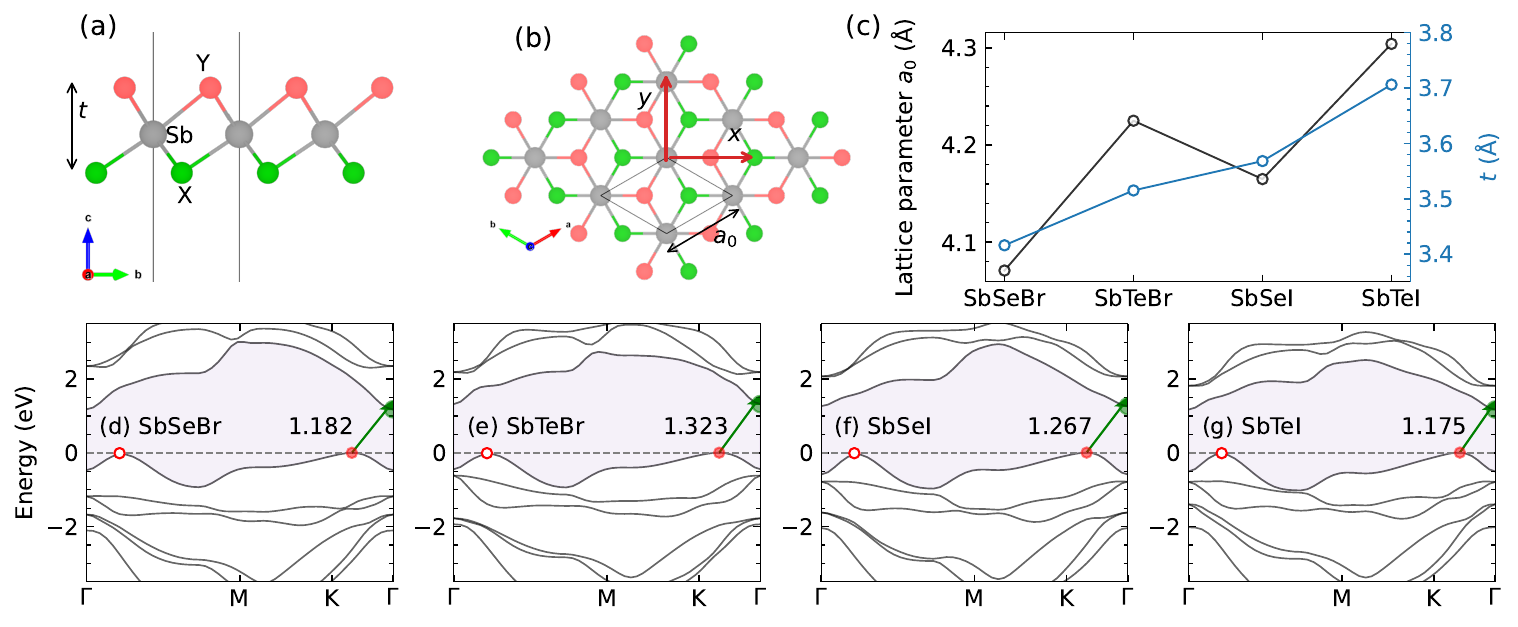}
 \caption{ The (a) cross-sectional and (b) perpendicular views of the sandwich atomic arrangement in SbXY JLs. The X and Y represent the chalcogen (Se or Te) and halide (Br or I) atoms, respectively. The vertical distance between X and Y is taken as the thickness $t$ of JLs, and the calculated $t$ is given in (c) along with the unitcell lattice parameter $a_0$. The energy levels of JLs in (d-g) are derived from the wavefunction coefficients calculated using the BoltzTrap code~\cite{btp2}. The Fermi level is set at $\SI{0}{\electronvolt}$, represented by the horizontal dashed gray line. The shaded region indicates all possible band gaps between valence and conduction bands. The shift in the valence band maximum (occurs along the $K-\Gamma$ direction) and conduction band minimum (positions at $\Gamma$ point) turns all SbXY JLs into indirect bandgap layers, denoted by green arrows with calculated bandgap (in \SI{}{\electronvolt}). The Cartesian X and $y$ directions are mentioned by arrows in (b).}
 \label{fgr:elebnd}
\end{figure*}

The ab initio molecular dynamics (AIMD) simulations have been carried out at \SI{300}{\kelvin} using the Quantum ESPRESSO package~\cite{Giannozzi_2009, Giannozzi_2017}, employing the Parrinello–Rahman extended Lagrangian-based cell dynamics. The Rappe–Rabe–Kaxiras–Joannopoulos ultrasoft pseudopotentials, from the Quantum ESPRESSO pseudopotential data base~\cite{qepseudo}, are used for Sb ($5s2~5p3$; \textit{Sb.pbe-n-rrkjus\_psl.1.0.0.UPF}), Br ($4s2~4p5$), I ($5s2~5p5$; \textit{I.pbe-n-rrkjus\_psl.1.0.0.UPF}), Se ($4s2~ 4p4$; \textit{Se.pbe-n-rrkjus\_psl.1.0.0.UPF}), and Te ($5s2 ~5p4$; \textit{Te.pbe-n-rrkjus\_psl.1.0.0.UPF})~\cite{AndreaDal2014}. The supercell of size $5a_0 \times 5a_0$, cf. Fig.~\ref{fgr:elebnd}(c) for lattice parameter $a_0$, is thermalized using the velocity rescaling method in the initial few \SI{}{\pico\second} of AIMD and further equilibrated at the targeted temperature with a time step of \SI{1.5}{\femto\second}. From the AIMD, the force-displacement dataset is generated and further processed using the Alamode code~\cite{TT2018_jpsj, TT2014_jpcm} to evaluate the phonon contribution to thermal conductivity as
\begin{equation}\label{eq:kappal}
\kappa _{l}^{\mu ,\nu} = \frac{1}{V N} \sum_{\boldsymbol{q}, j} c\left( \boldsymbol{q}, j \right) v^{\mu}\left(\boldsymbol{q}, j\right) v^{\nu}\left(\boldsymbol{q}, j\right) \tau \left(\boldsymbol{q}, j\right),
\end{equation}
where $V$, $n$, $c\left(\boldsymbol{q}, j\right)$,  $v\left(\boldsymbol{q}, j\right)$, $\tau \left(\boldsymbol{q}, j\right)$ are the volume of the simulation cell, and the number of $q$ points in the first Brillouin zone, the constant-volume specific heat, the phonon group velocity, and the phonon relaxation time, respectively. The indices $\mu$ and $\nu$ represent Cartesian components of the indexed quantities.

\begin{figure*}[tb]
 \centering
 \includegraphics[width=1\textwidth]{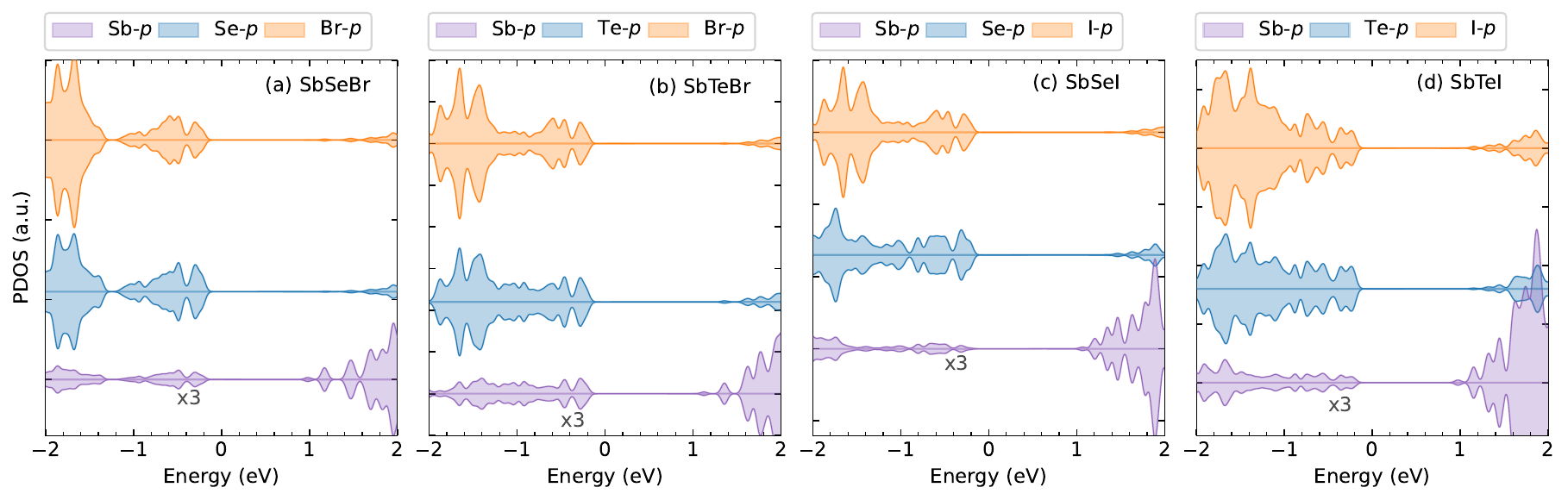}
 \caption{Computed partial density of states in arbitrary units. The valence bands are contributed by $p$ orbitals of both the chalcogen X and halide Y in all layers, whereas Sb-$p$ orbitals dominate the conduction band. The equal up and down spin states depict that the 1T-phases of SbXY JL are essentially non-magnetic. Note that the Sb-$p$ states are magnified by a factor of three for better visibility.}
 \label{fgr:dos}
\end{figure*}

\section{\label{sec:rd}Results and Discussions}
\subsection{\label{sec:elecpro}Electronic properties}
The SbXY JLs are modeled by sandwiching the Sb cationic atomic layer in between the anionic atomic layers of chalcogen X (Se or Te) and halide Y (Br or I), in a trigonal conventional cell with space group symmetry \texttt{Pm31} (no. 156), cf. Fig.~\ref{fgr:elebnd}(a-b). The optimized lattice parameter of the layer by in-plane structural relaxation varies from \SIrange{4.07}{4.30}{\angstrom}, originating from the difference in the atomic size of X and Y. Like the lattice parameter, the thickness $t$ of the layers varies on a small scale from \SIrange{3.42}{3.71}{\angstrom}, cf. Fig.~\ref{fgr:elebnd}(c). As the atomic sizes of X and Y increase, the steric effect in SbXY JLs induces a linear trend on $t$ such that exposing the $t_{\rm{SbSeBr}}<t_{\rm{SbTeBr}}<t_{\rm{SbSeI}}<t_{\rm SbTeI}$ correlation in which the atomic size of either X or Y always increases. The observed $t$ and lattice parameters are in the order of previous reports~\cite{poonam2024jmca, Bafekry2021, SDGuo2017, ASE2024PCCP}. The chalcogen and halide combination varies the $t$ in a range of $\approx \SI{0.30}{\angstrom}$ and lattice parameter by $\approx \SI{0.23}{\angstrom}$. Therefore, the JLs are expected to have similar electron distribution along high-symmetric directions in each layer, though the bonding character is altered by the presence of different X and Y. 

The calculated electric band structure, see Fig.~\ref{fgr:elebnd}(d-g), reveals that all SbXY JLs are indirect bandgap semiconductors, with the bandgap varying from \SIrange{1.175}{1.323}{\electronvolt}. They exhibit a feature that the last few valence bands near the Fermi level (shifted and fixed at $\SI{0}{\electronvolt}$) are isolated from remaining valence energy states, particularly the presence of a distinct valence band in the SbSeBr JL. The overlap between valence energy states increases as the atomic size of either X or Y or both X-Y increases. Additionally, the calculated partial density of states, shown in Fig.~\ref{fgr:dos}, corroborates that the orbital characteristics of each valence and conduction bands originate from the $p$-orbitals of constituent atoms. The $p$-orbitals of X and Y predominate in the valence bands, while the $p$-orbitals of Sb, along with a minor portion of $p_z$-orbitals of Y, contribute to the conduction bands, which is in good agreement with previous reports~\cite{Bafekry2021, ASE2024PCCP}. Note that, when the atomic size of X and Y increases, the hybridization between Sb-$p$ and $p_z$-orbitals of both X and Y also increases, in agreement with the valence energy states shown in Fig.~\ref{fgr:elebnd}(d-g). Since $p$-orbitals of X and Y exhibit similar characteristics, cf. Fig.~\ref{fgr:dos}, the electronic bands invariably show a nearly similar profile, cf. Fig.~\ref{fgr:elebnd}(d-g), particularly at the valence band maximum. 

Apart from the isolated band, the valence band exhibits a secondary maximum (cf. open circles in Fig.~\ref{fgr:elebnd}(d-g)), in addition to the valence band maximum (cf. filled circles in Fig.~\ref{fgr:elebnd}(d-g)), which originates due to the fact of similar energies of $p$-orbitals of X and Y. These secondary maxima differ from the valence band maximum by a few $\SI{}{\milli\electronvolt}$, but are oriented along $\Gamma-M$ high-symmetric direction. The calculated energy differences in SbSeBr, SbTeBr, SbSeI, and SbTeI are $\SI{45}{\milli\electronvolt}$, $\SI{39}{\milli\electronvolt}$, $\SI{60}{\milli\electronvolt}$, and $\SI{39}{\milli\electronvolt}$, respectively. The primary and secondary maxima in the valence band present hole carrier pockets, which may improve their mobility. Similarly, the electron carrier pocket is observed in the conduction band only at the $\Gamma$ point, where minima occur. However, other than carrier pockets, band curvature influences the carrier characteristics dominantly and determines the effective mass of charge carriers, thereby controlling their mobility.

\begin{figure*}[tb]
 \centering
 \includegraphics[width=1\textwidth, clip, trim=0.9in 0.32in 0.28in 1in]{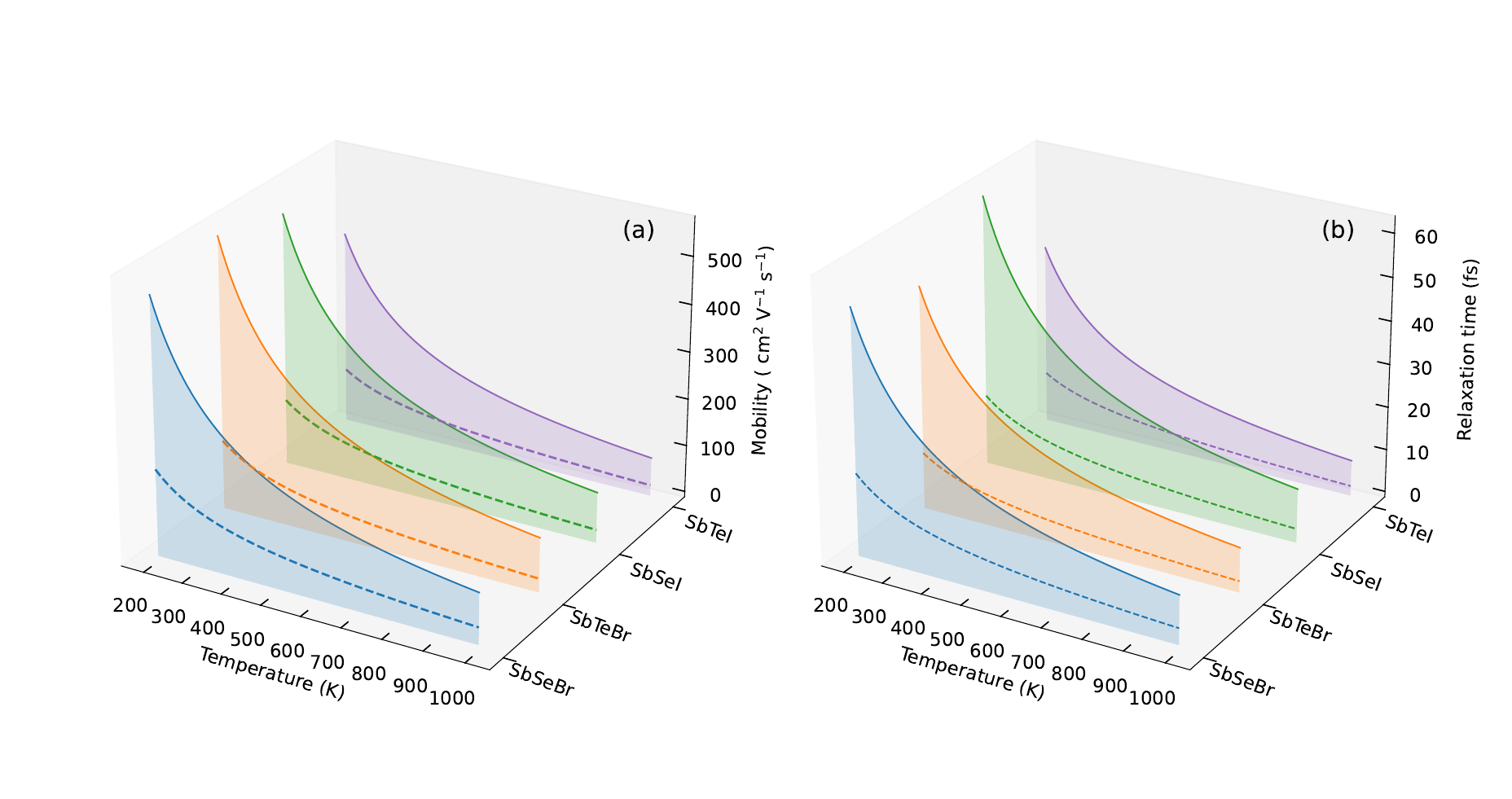}
 \caption{Summarized are the calculated (a) mobility $\mu$ and (b) relaxation time $\tau_c$ of electrons in SbXY JLs. The $x$- and $y$-components of the presented quantities are denoted by continuous and dashed lines, respectively. The mobility of electrons is higher than that of holes by a factor of $\approx \SI{10}{}$, cf. Fig. S1 in SM. Consequently, holes have a shorter mean free time $\tau_c$ and relax faster. The anisotropies in $\mu$ and $\tau_c$ denote that the SbXY JLs may have different TE efficiency in different directions.}
 \label{fgr:e_mu-tau}
\end{figure*}
\begin{figure*}[tb]
 \centering
 \includegraphics[width=0.9\textwidth]{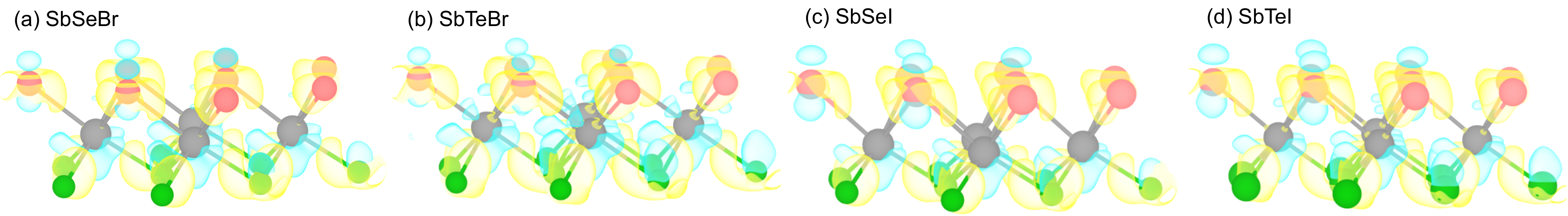}
 \caption{Computed differential charge density $\Delta \rho$ in SbXY JLs. The yellow (cyan) isosurfaces, visualized at \SI{4e-2}{} $e$/bohr$^3$ level, denote charge accumulation (depletion from) on atoms. The charge depletion from Sb indicates the cationic character, and accumulation on X and Y denotes their anionic character. The clear separation between charge depletion and accumulation results in bonding that is mostly ionic in nature. The qualitative assessment reveals that the charge distribution in SbTeBr JL extends over a larger area compared to other JLs, indicating improved charge dynamics.}
 \label{fgr:cdd}
\end{figure*}

\subsection{\label{sec:mutau}Mobility and relaxation time of carriers}
The carrier mobility $\mu$ is investigated under the deformation potential (DP) approximation~\cite{Shuai2012, Bardeen1950, Xi2012}. According to the DP approximation, the carrier relaxation time $\tau_c$ is calculated from the relation $\tau_c \, = \, \mu m^* / e $, where $\mu$, $m^*$, and $e$ are mobility, effective mass of charge carrier, and electron charge, respectively. The relationship between $\mu$ and other characteristics of layers is given in Sec. 2 in SM. The obtained variations of $\mu$ and $\tau_c$ as a function of temperature for electron carriers are given in Fig.~\ref{fgr:e_mu-tau} and for hole carriers in Fig. S1 in SM. In general, the electron carriers are more mobile than the hole carriers due to the curvature of bands, resulting in lighter electrons and heavier holes. The effective masses of electrons and holes are analyzed using different methods, such as band fitting and differentiation methods, whose results are summarized in Sec. 3 in SM. The effective masses obtained from the effective mass tensor are utilized in further calculations. The effective mass tensor is calculated using differentiation methods as implemented in the \textit{emc} code~\cite{emc} and using fitting methods as implemented in the \textit{effmass} code~\cite{Whalley2018}. The parameters used to calculate $\mu$ and effective masses of carrier are summarized in Tab.~\ref{tab:tau_elec} 

Comparing $\mu$ in Fig.~\ref{fgr:e_mu-tau}(a) of JLs shows that the electron carriers in SbTeBr JLs are dynamic than those in the remaining layers, and their $\mu$ reaches as high as \SI{380}{\centi\meter\squared\per\volt\per\second} at \SI{300}{\kelvin}, the highest among SbXY JLs. The $\mu$ of electrons in SbSeBr and SbSeI JLs has a nearly similar trend across the considered temperature range and is smaller than that in the SbTeBr layer. The smallest $\mu$ is observed for the SbTeI JL. Our results for $\mu$ deviate by approximately an order of magnitude from the literature value (\SI{3810}{\centi\meter\squared\per\volt\per\second} at \SI{300}{\kelvin}) \cite{poonam2024jmca}, which appears dubious. Obviously, the difference from the literature data is due to the unclear specification of the carrier type and effective mass, as well as the unusually low reported relaxation time (\SI{0.003}{\pico\second}). The shorter $\tau_c$ denotes stronger scattering of carriers and poor conductivity. Our results suggest that the electrons in SbTeI undergo frequent scattering, resulting in the least observed $\tau_c$ among SbXY JLs. Except for SbTeI, the $\tau_c$ of electrons in the remaining layers has similar values from 100 K to 1000 K, reaching the maximum of 41 fs at \SI{300}{\kelvin} in the SbSeI JL. 

Since $\mu$ involves the elastic nature of the materials, the carriers in the layers with a lower elastic modulus $C_{\beta}$ show lower $\mu$ and scatter frequently. The anisotropic $C_{\beta}$ of SbXY JLs is correlated with the directional dependency of $\mu$, exhibiting the $\mu^y$ (mobility along the $y$ direction) always lower than that along the $x$ direction. Therefore, it is expected that only the $x$-directional moment of electron carriers has a considerable influence on determining the TE properties of SbXY JLs. The holes in the valence band attain a relatively higher effective mass than electrons due to less curvature of the valence band in comparison to the conduction band, cf. Tab.~\ref{tab:tau_elec}. Hence, the $\mu$ and $\tau_c$ of holes are smaller than those of electrons, along both $x$- and $y$-directions. 

To understand the driving force governing variation in $\mu$ of electrons and holes, the charge redistribution $\Delta \rho$ between the cationic Sb$^+$ and the anionic $X^-$/$Y^-$ is examined, see Fig.~\ref{fgr:cdd}, which reveals a clear charge separation, indicating predominant ionic character in Sb–X and Sb–Y bondings. Additionally, the shape and orientation of the charge isosurfaces suggest that the bonding originates predominantly from the $p$ orbitals of the constituent atoms. The calculated ionicity, using the electronegativity difference between Sb-$X/Y$ of the bonds Sb-Se (\SI{6.06}{\percent}), Sb-Te (\SI{0.06}{\percent}),  Sb-Br (\SI{18.70}{\percent}), and Sb-I (\SI{8.90}{\percent}) in SbXY layers shows that the Sb–Te bond has the most covalent character, while Sb–Br exhibits the highest ionicity. Such a unique combination of covalent Sb–Te and ionic Sb–Br bonding in SbTeBr leads to a balanced electronic condition, which enhances carrier delocalization and suppresses carrier scattering, resulting in the observed high $\mu$. 

\begin{table*}[tb]
\small
  \caption{The calculated elastic modulus $C_{\beta}$, deformation potential constant $E_d$, electron effective mass $m^*$, and hole effective mass $m^*$. The effective masses of electron and hole are given in units of electron rest mass $m_e$.}
  \label{tab:tau_elec}
  \begin{tabular*}{1\textwidth}{@{\extracolsep{\fill}}lccccc}
    \hline
    \hline
    Layer   &  Direction & $C_{\beta}$ (Nm$^{-1}$) & $E_d$ (eV) & Electron $m^*$  & Hole $m^*$   \\
    \hline
    SbSeBr   & $x$   & 34.14  &   7.73  &   0.184   &   0.735  \\
             & $y$   & 10.40  &   7.33  &   0.184   &   0.735  \\
    SbTeBr   & $x$   & 31.58  &   6.85  &   0.206   &   0.698  \\
             & $y$   &  7.82  &   7.83  &   0.206   &   0.777  \\
    SbSeI    & $x$   & 33.36  &   8.41  &   0.158   &   0.523  \\
             & $y$   & 11.22  &   8.31  &   0.158   &   0.648  \\
    SbTeI    & $x$   & 30.44  &   8.61  &   0.179   &   0.680  \\
             & $y$   &  8.32  &   8.57  &   0.179   &   0.534  \\
    \hline
    \hline
  \end{tabular*}
\end{table*}
The 1T phase of SbXY exhibits an inherently anisotropic atomic arrangement along the $x$- and $y$- directions, see direction notations in Fig.~\ref{fgr:elebnd}b, leading to distinct potential energy surfaces under applied strain. Such anisotropy manifests in the directional dependence of $\mu$, which originates from variations in the deformation potential and consequently modulates both the elastic modulus and the effective mass of charge carriers, see Tab.~\ref{tab:tau_elec}. Owing to the relation $\mu \propto C_{\beta}$, a threefold difference in modulus between the $x$ and $y$ directions results in consistently higher electron mobility along the $x$-direction. This anisotropic $\mu$ is also reflected in the relaxation time $\tau_c$ of electrons, see Fig.~\ref{fgr:e_mu-tau}(b). As compensated by the effective mass of electrons, the $\tau_c$ of electrons in SbSeBr, SbTeBr, and SbSeI JLs attain similar values for individual temperatures across the considered temperature range.

Consequently, holes in the valence bands exhibit stronger localization, primarily due to the presence of more electronegative X and Y anions, enhancing the polar scattering. As a result, the $\mu$ of holes is significantly reduced. Notably, the holes in SbSeI JL exhibit the lowest observed $\mu$ (see Fig. S1 in SM). Furthermore, the $\tau_c$ for holes is nearly half that of electrons, indicating that holes are scattered more than electrons. Direct comparison of $\mu$ of electrons and holes reveals that electron mobility exceeds that of holes by approximately a factor of 10.

The interplay between bonding polarity and charge redistribution governs the asymmetry between electron and hole mobilities, highlighting the potential for significant differences in thermoelectric efficiency. The Te-Br bonding analysis in SbTeBr reveals the origin of its superior electron transport properties. These results suggest that SbXY JLs can exhibit enhanced thermoelectric performance at both room and elevated temperatures. Additionally, the disparity between the behavior of electrons and holes may mitigate the bipolar effect, and the concentration of electrons can be selectively enhanced to achieve the maximum possible TE efficiency in SbXY JLs.

\subsection{\label{sec:phonon}Thermal stability of SbXY layers}
\begin{figure*}[tb]
 \centering
 \includegraphics[width=1\textwidth]{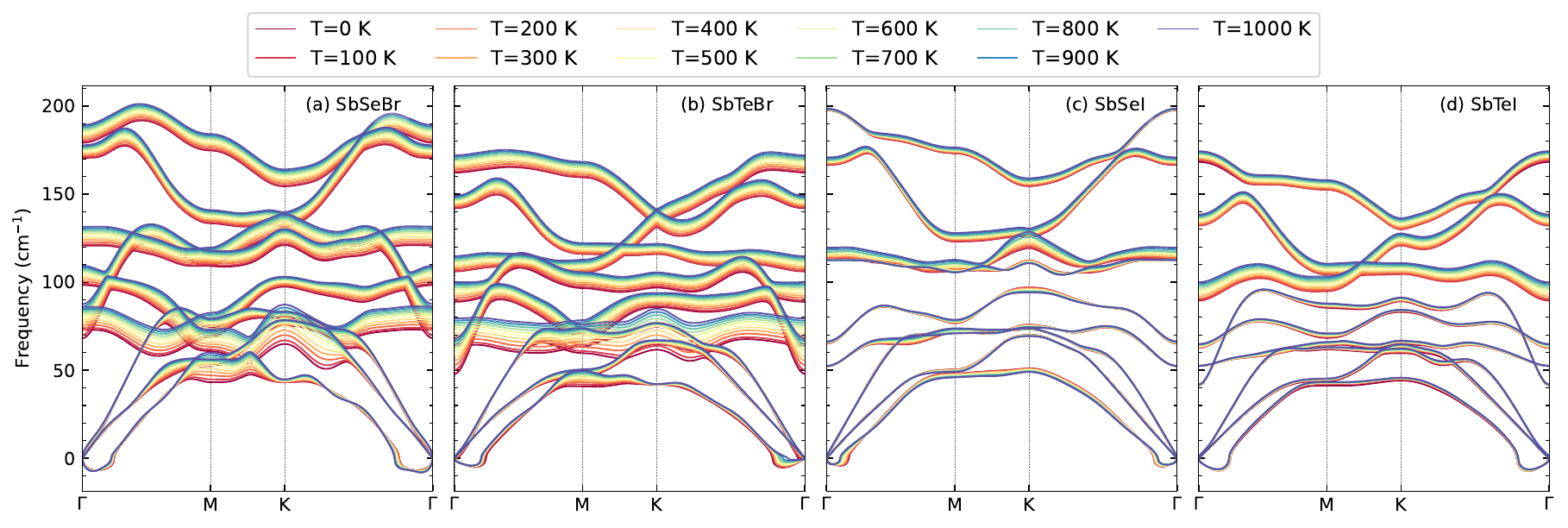}
     \caption{Temperature-dependent phonon dispersions of SbXY JLs from 0 K – 1000 K. Acoustic branches remain largely rigid with only minor variation at high-symmetry points, whereas the optical modes display temperature-driven hardening. In SbSeI, the presence of I$^-$ suppresses anharmonic effects, yielding a comparatively rigid vibrational spectrum.}
 \label{fgr:scph}
\end{figure*}

The SbXY JLs have been demonstrated to be stable within the DFT accuracy~\cite{ASE2024PCCP, Bafekry2021, SDGuo2017}. However, their stability at elevated temperatures remains an open question. To address this, self-consistent phonon (SCP) calculations are employed to obtain temperature-dependent phonon dispersions. The SCP method accounts for anharmonic effects by renormalization of phonon frequencies at finite temperatures. As implemented in the Alamode \cite{TT2014_jpcm, TT2018_jpsj}, the SCP approach solves the Dyson equation within the many-body Green function formalism, using both harmonic and anharmonic interatomic force constants (IFCs) to accurately describe the phonon many-body problem. The renormalization of phonon frequencies at elevated temperatures is determined by minimizing the function \cite{Tadano2015},
\begin{equation}   
\Omega_q^{2}=\omega_{q}^{2} + \Omega_q \sum_{q_1} \frac{\hbar \Phi(q;-q;q_1;-q_1)}{4\Omega_q \Omega_{q_1}} \left[1 + 2n(\Omega_{q_1})\right], 
\label{eq:omegaq}
\end{equation}
where, $\Omega_{q}$ denotes the anharmonic phonon frequency at finite temperature, $\omega_{q}$ is the phonon frequency within the harmonic approximation, $\Phi(q; -q; q_1; -q_1)$ represents the phonon self-energy arising from the quartic anharmonicity, and $n(\Omega_{q_1})$ is the Bose–Einstein distribution function. To solve Eq. \ref{eq:omegaq}, both quartic and cubic IFCs are extracted from a force-displacement dataset generated via AIMD simulations, see Sec. \ref{sec:method}. At each AIMD step, the atomic forces and corresponding displacements are recorded, forming the input for IFC fitting. The least absolute shrinkage and selection operator (LASSO) regression technique is employed to efficiently solve for $\tilde{{\Phi}}$~\cite{Tadano2015, Zhou_PRL2014},
\begin{equation}
    \tilde{{\Phi}} = argmin_{\Phi} \parallel A {\Phi}-\mathbf{F} \parallel_2^2 + \lambda \parallel {\Phi}\parallel_1 ,
    \label{er:phi}
\end{equation}
where ${\Phi}$ is the parameter vector consisting of linearly independent IFCs, $\mathbf{F}$ is the vector of atomic forces derived from AIMD, $A$ is the matrix that contains atomic displacements, and $\lambda\parallel {\Phi}\parallel_1$ is the  penalty term. By Eqs.~\ref{eq:omegaq} and \ref{er:phi}, the phonon spectrum at elevated temperature is determined, see Fig.~\ref{fgr:scph}. 

The SCP temperature-dependent phonon calculations (0 K – 1000 K) for SbXY JLs (each has 9 branches, splitting into 3 acoustic and 6 optical branches) reveal distinct anharmonic behavior that correlates with mass and bonding. In SbSeBr and SbTeBr JLs, the optical bands show noticeable thermal renormalization, while the acoustic branches are largely inert except near the zone-boundary points $M$ and $K$, indicating that the phonon anharmonic coupling is strongest for finite $q$ modes, \ie near the zone boundary. The SbSeI and SbTeI layers exhibit stronger mode selectivity, with the lowest acoustic and first three optical branches essentially stable, likely due to the presence of the heavier anion I$^-$ and weak anharmonic low-frequency modes. The last three optical modes shift with temperature, particularly more pronounced in SbTeI, suggesting that the Sb-I and Se-Te/Se bonds are sensitive to mode-specific vibrations. The temperature-independence branches in SbSeI show stiffer effective bonding and suppressed anharmonic renormalization, which may result in higher thermal conductivity. 

The mean-square displacements (MSD) are computed to analyze the vibrational behavior of each atom, and the component-resolved MSD is shown in Fig. S6 in SM. The MSD increases monotonically with temperature from 0 K to 1000 K, reflecting thermally activated vibrations. Their magnitudes are element-dependent, governed by the interplay between atomic mass and local bonding strength. The Se atoms exhibit consistently smaller displacements along the $x$, $y$, and $z$ directions compared with the constituent chalcogens and halogens, indicating that the Sb atomic plane is more bound. Owing to its lighter mass relative to I, Br shows significantly larger oscillation amplitudes, which induce strong anharmonicity in the SbSeBr and SbTeBr layers. These enhanced vibrational amplitudes suggest that the pronounced anharmonicity in the Br-containing systems enables phonon scattering, thereby reducing thermal conductivity and enhancing TE performance compared with the I-containing layers. Additionally, the intact phonon spectrum indicates the structural stability of the assumed 1T phase of SbXY up to 1000 K. 

\subsection{\label{sec:tc}Phonon thermal conductivity}
\begin{figure*}[tb]
 \centering
 \includegraphics[width=0.9\textwidth, clip, trim=1.7in 0.5in 0.7in 1.1in]{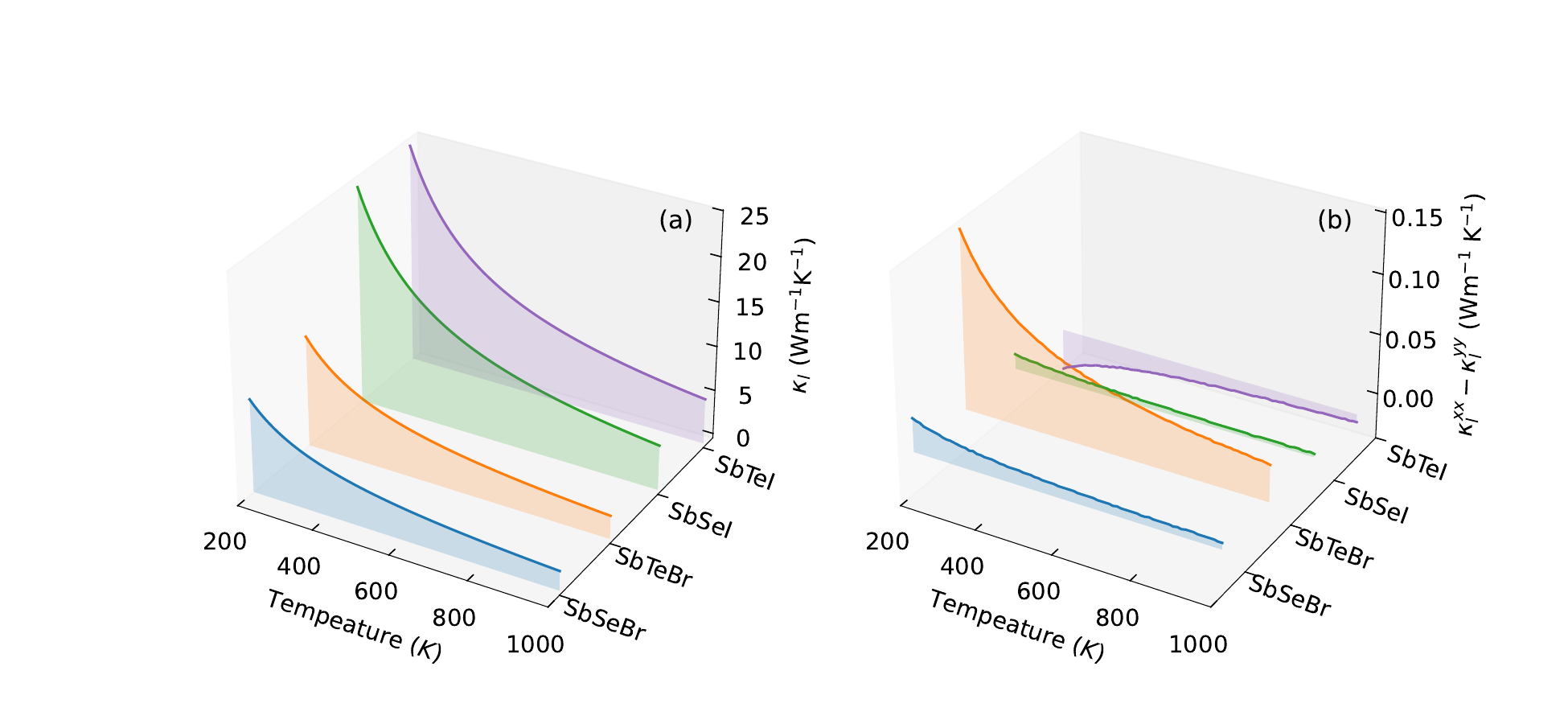}
 \caption{The calculated (a) lattice thermal conductivity of SbXY JLs as a function of temperature. The anisotropy between $\kappa_l$ components, defined as $\Delta \kappa_l ~=~\kappa_l^{xx}-\kappa_l^{yy}$, is shown in (b).}
 \label{fgr:kl}
\end{figure*}
Based on the force-displacement dataset from AIMD, the phonon–phonon interaction is analyzed to estimate the lattice thermal conductivity $\kappa_l$ using the anharmonic lattice dynamics (ALD) method~\cite{TT2014_jpcm} which utilizes Boltzmann transport theory~\cite{TT2014_jpcm, kaur2019, Siddi2024}. As a standard practice, the computed $\kappa_l$ is scaled up by the factor $z/t$, where $z$ is the length of the unitcell along crystallographic direction $c$ and $t$ is the thickness of the JLs~\cite{WU2017233, Ould2024}. The calculated total $\kappa_l$ (\ie average of components) and the anisotropy in $\kappa_l$ components $\Delta \kappa_l ~=~\kappa_l^{xx}-\kappa_l^{yy}$ are shown in Fig.~\ref{fgr:kl}. 

The comparison of $\kappa_l$ of SbXY JLs depicts that the Br-containing JLs exhibit lower $\kappa_l$ than the counterpart JLs containing I. Though I is heavier than Br, the charge distribution on Sb-I and Sb-X, and the electronegativity difference between Sb, X, and I result in lower ionic bonding than that of Sb-Br bonding, see Sec.~\ref{sec:elecpro}. Ionic bonding tends to have a large charge separation between constituent atoms, which enhances long-range Coulomb interactions. Such interactions cause optical phonons, with more polar characteristics, to couple strongly with acoustic phonons via Fr\"ohlich interaction. Therefore, the strong Sb-Br ionic bonding in SbSeBr and SeTeBr JLs reduces the thermal conductivity through phonon-phonon scattering, causing the phonon to be short-lived. In all SbXY JLs, $\kappa_l$ decreases with increasing temperature, consistent with enhanced phonon-phonon scattering at higher temperatures and falls off as $T^{-1}$. Such behavior is similar to the observations for other 2D layers, such as MX ($M$ = Ge, Sn; X = S, Se)~\cite{WU2017233,Qin2016}, Pb$_2$Sb$_2$X$_5$ (X = S, Se, Te)~\cite{YGan2021} and PbXY ($X,~Y$= F, Cl, Br, I) layers~\cite{VM2025}.

At 300 K, the calculated $\kappa_l$ values of SbXY JLs are 7.26 \klunit{} (SbSeBr), 8.68 \klunit{} (SbTeBr), 16.90 \klunit{} (SbSeI), and 17.08 \klunit{} (SbTeI), indicating that the thermal conductivity of Br-containing layers is approximately half that of the I-containing counterparts. The present $\kappa_l$ value for the SbSeI JL (16.90 \klunit{}) is in good agreement with the previously reported value of 14.2 \klunit{} at 300 K~\cite{Jdong2024}. The non-normalized (by $z/t$) $\kappa_l$ values of SbXY JLs are comparable to other 2D layers that contain X-Y combination of anions, such as AsTeBr (2.02 \klunit{}) \cite{Poonam2023}, AsTeI (3.36 \klunit{}) \cite{Poonam2023}, BiTeBr (1.47 \klunit{}) \cite{GUO2017}, and SbTeI (3.55 \klunit{}) \cite{Chu2023}. The anisotropy $\Delta \kappa_l$, defined as the difference between the $\kappa_l^{xx}$ and $\kappa_l^{yy}$ components, is negligible across the studied temperature range for SbXY JLs, see Fig.~\ref{fgr:kl}(b). Notably, although $\Delta \kappa_l$ is more pronounced in the SeTeBr JL, the relative anisotropy $\Delta \kappa_l / \kappa_l$   remains within 0.62\%–0.59\% over the range 200 K – 1000 K. Therefore, $\Delta \kappa_l$ is not considered in the subsequent estimation of the thermoelectric properties of SbXY JLs.

\begin{figure*}[tb]
 \centering
 \includegraphics[width=1\textwidth, clip, trim=0in 0in 0in 0in]{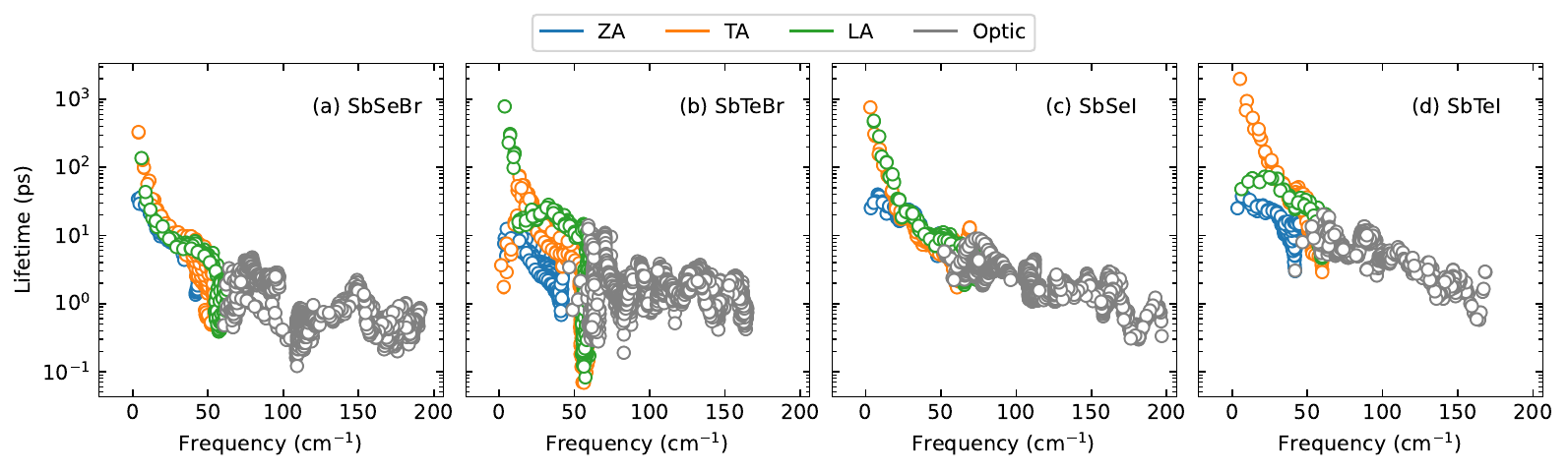}
 \caption{The mode-resolved phonon lifetime $\tau\left(\boldsymbol{q}, j\right)$ as a function of frequency in SbXY JLs, derived using the ALD method. The optical phonon modes have a shorter life span than the acoustic modes. Also, the lifetime in Br-containing layers (SbSeBr and SbTeBr) is shorter than that in I-containing layers (SbSeI and SbTeI).}
 \label{fgr:phonon-lt}
\end{figure*}

To understand the variation in $\kappa_l$ of SbXY JLs, the lifetime  $\tau\left(\boldsymbol{q}, j\right)$ of phonons is analyzed, where $\boldsymbol{q}$ and $j$ are the wavevector and phonon mode, respectively. Note that the curvature on phonon bands, see Fig.~\ref{fgr:scph}, hints at the heat-carrying dynamics as individual phonon modes contribute to $\kappa_l$. Hence, the phonon with a longer lifetime continuously carries the heat, whereas phonons with a short lifetime dissipate shortly and hinder the heat-carrying process~\cite{MinLi2024prb}. The calculated $\tau\left(\boldsymbol{q}, j\right)$ is shown in Fig.~\ref{fgr:phonon-lt}, and it is observed that the acoustic modes are more stable than the optical modes in all SbXY layers. Additionally, the optical modes in I-containing layers are sustained longer than those in Br-containing layers, indicating that the contribution to $\kappa_l$ optical modes enhances the thermal conductivity in I-containing layers. In particular, transverse and longitudinal acoustic phonons dominate the thermal transport of ZA modes. The $\tau\left(\boldsymbol{q}, j\right)$ of phonon modes demonstrates that the longer sustaining optical phonons cause higher $\kappa_l$ in SbSeI and SbTeI layers.

\subsection{\label{sec:TEprop}Thermoelectric properties}
Based on the linearized Boltzmann transport equation, the TE properties are computed, imposing the constant relaxation time approximation. It is standard that the transport properties, electrical conductivity, and electronic thermal conductivity obtained from BoltzTrap are normalized by carrier relaxation time $\tau_c$. Therefore, we calculated the $\tau_c$ using deformation potential theory, a standard approach for 2D materials in the absence of experimental results~\cite{Mamani2024, YGan2021, Shafique2017, MILI2022cms}, for electrons and holes at each assumed temperature to get $\sigma$ and $\kappa$. As shown in Fig.~\ref{fgr:e_mu-tau}, the anisotropy in mobility $\mu$ suggests a potential difference in TE performance along $x$ and $y$ directions. Due to this fact, the power factor $PF$ and Figure of Merit $ZT$ are computed separately for each direction using the corresponding $\tau_c$ values for both electrons and holes. 

\begin{figure*}[tb]
 \centering
 \includegraphics[width=1.0\textwidth, clip, trim=0in 0.32in 0in 0in]{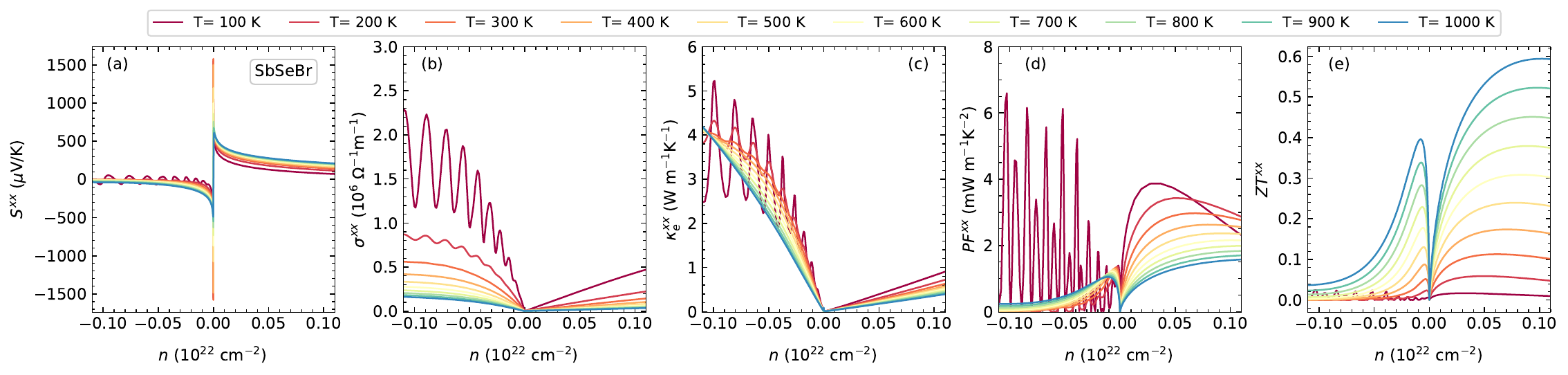}
 \includegraphics[width=1.0\textwidth, clip, trim=0in 0in 0in 0.4in]{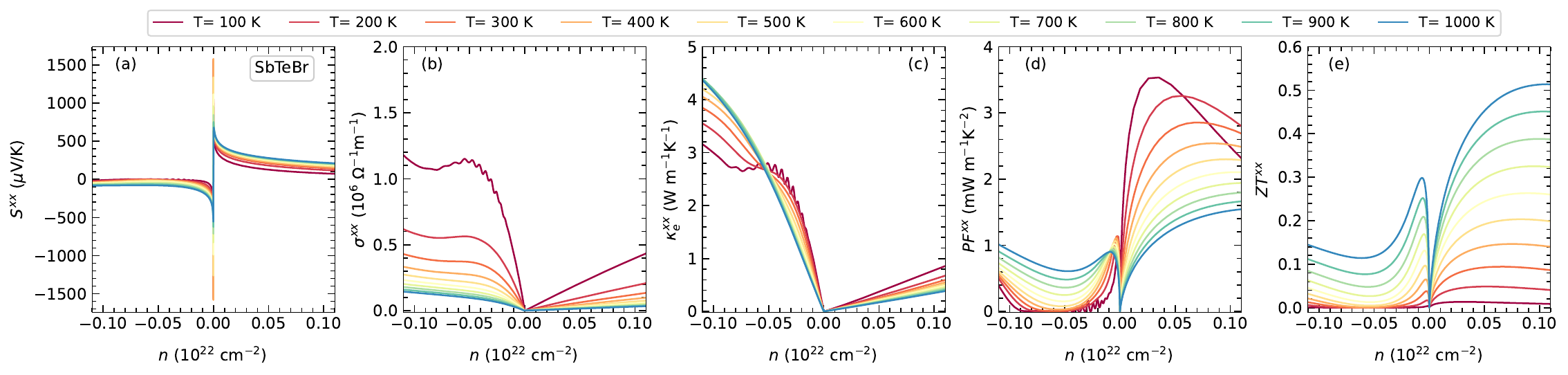}
 \caption{The calculated $xx$-components of a) Seebeck coefficient $S$, b) electrical conductivity $\sigma$, c) electronic thermal conductivity $\kappa_e$, d) power factor $PF$, and e) Figure of Merit $ZT$ of Br-containing layers SbSeBr (top row) and SbTeBr (bottom row). The positive (negative) carrier concentration $n$ denotes hole (electron) doping by shifting the Fermi level towards the valence (conduction) band edge.}
 \label{fgr:zt-br}
\end{figure*}
\begin{figure*}[tb]
 \centering
 \includegraphics[width=1\textwidth, clip, trim=0in 0.3in 0in 0in]{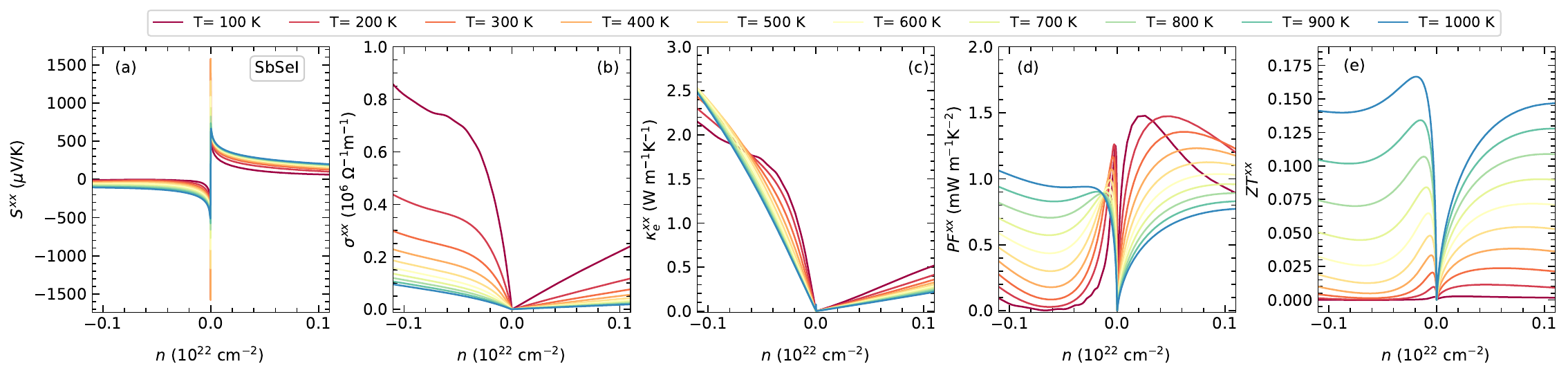}
 \includegraphics[width=1\textwidth, clip, trim=0in 0in 0in 0.4in]{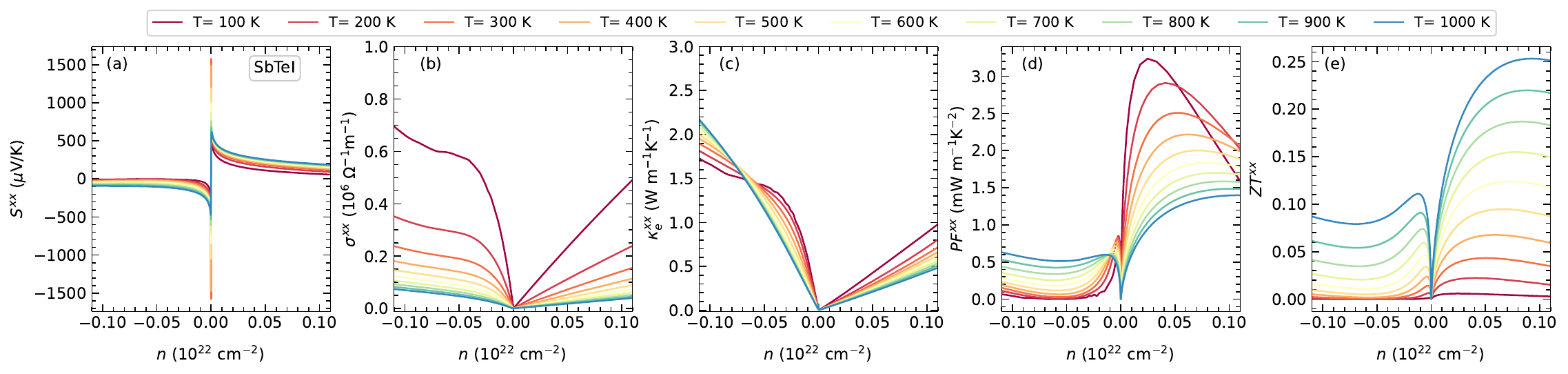}
 \caption{The calculated $xx$-components of a) Seebeck coefficient $S$, b) electrical conductivity $\sigma$, c) electronic thermal conductivity $\kappa_e$, d) power factor $PF$, and e) Figure of Merit $ZT$ of I-containing layers SbSeI (top row) and SbTeI (bottom row). The positive (negative) carrier concentration $n$ denotes hole (electron) doping by shifting the Fermi level towards the valence (conduction) band edge.}
 \label{fgr:zt-i}
\end{figure*}
The $xx$-components of TE parameters Seebeck coefficient $S$, electrical conductivity $\sigma$, electronic thermal conductivity $\kappa_e$, and derived power factor $PF~(=S^2\sigma)$, are presented as a function of carrier concentration $n$ in Figs.~\ref{fgr:zt-br} and \ref{fgr:zt-i} for Br- and I-containing layers, respectively. The $yy$-component of these parameters is given in Fig. S7-S8 in SM. Note that the relaxation times of electrons and holes are applied if $n<0$ and $n>0$, respectively. The carrier concentration is calculated by fixing the Fermi level at the midpoint between the conduction band minimum and the valence band maximum, following the convention implemented in BoltzTrap within the rigid band approximation framework \cite{TS2003, GM2006}. Therefore, the negative (positive) values of $n$ correspond to electron (hole) doping, obtained by shifting the Fermi level toward the conduction (valence) bands.

The bandgap of SbXY semiconductors opens up the opportunity to use them for TE applications, where the TE efficiency is enhanced through tuning carrier concentration. However, the effective mass of carriers increases with increasing carrier concentration. Though the heavier charge carriers improve the Seebeck coefficient, their mass reduces the electrical conductivity due to lower mobility~\cite{snyder08_natmat}. Hence, identifying an optimal carrier concentration that balances these contrasting effects is crucial for maximizing the Figure of Merit $ZT$ and assessing the TE potential of these materials. For practical applications, it is particularly advantageous if the optimal carrier concentration is closer to the Fermi level~\cite{Snyder_2012_EnergyEnvironSci, PJin2022}.

In general, the TE performance in the $xx$ direction is higher than that along the $yy$ direction due to lower effective mass, higher mobility, and enhanced scattering, cf. Fig.~\ref{fgr:e_mu-tau} and Fig. S1 in SM. Since the Seebeck coefficient is proportional to the bandgap of the material~\cite{gibbs2015}, the maximum $|S|$ values along the $xx$ direction in SbXY layers are similar. Also, these maximum values of $|S|$ are observed within a narrow carrier concentration range close to the Fermi level, cf. column (a) in Fig. \ref{fgr:zt-br} - \ref{fgr:zt-i}. The summarized maximum $|S|$ of  SbXY, cf. Fig. S9(a), shows that a maximum of approximately $\approx \SI{1500}{\micro\volt\per\kelvin}$ is reached at $\SI{300}{\kelvin}$, comparable to Pb-based 2D layers  Pb$_2$SSe \cite{Mamani2024} and PbSnS$_2$ \cite{CDing2023} and potentially lower than that of MoS$_2$ \cite{MV2020}, Si$_2$PH$_2$ and Ge$_2$PH$_2$ \cite{Mohebpour2021}. Since $|S|$ is not affected by the relaxation time $\tau_c$ of carriers, a similar behavior is also observed along the $yy$ direction, cf. Fig S7-S8.

The $\tau_c$-dependent quantity, electrical conductivity $\sigma$, is a key factor in determining the thermoelectric performance of the materials, as higher carrier mobility leads to larger $\sigma$. The calculated $xx$ component of $\sigma$ is shown in Fig. \ref{fgr:zt-br}(b)–\ref{fgr:zt-i}(b), while the $yy$ component is provided in Fig. S7-S8. The comparison reveals that $\sigma$ due to electrons is consistently higher than that from holes, owing to the lower effective mass of electrons relative to holes. Among the SbXY layers, SbSeBr exhibits the highest $\sigma$ (cf. Fig. \ref{fgr:zt-br}(b)). At 300 K, the $\sigma$ due to electrons reach the maximum of $\SI{0.6E6}{}$ \sgunit{} for SbSeBr and $\SI{0.4E6}{}$ \sgunit{} for SbTeBr. In contrast, I-containing compounds exhibit reduced values of $\SI{0.3E6}{}$ \sgunit{} (SbSeI) and $\SI{0.2E6}{}$ \sgunit{} (SbTeI). The $\sigma$ of SbTeI is weakest due to the lowest carrier mobility for both electrons and holes. The enhanced $\sigma$ originates mainly from the lighter electrons, while heavier holes suffer from stronger scattering and contribute little. Furthermore, mobility along the $yy$ direction is smaller than along $xx$ (cf. continuous vs. dashed lines in Fig. \ref{fgr:e_mu-tau}), leading to reduced $\sigma_{yy}$. Such directional and carrier-selective $\sigma$ is favorable for thermoelectric performance, since the $\sigma$ dominated by a single carrier type subdues detrimental mixed conduction from both electrons and holes \cite{snyder08_natmat}.

Apart from the need for high $S$ and $\sigma$, achieving the electronic thermal conductivity $\kappa_e$ as low as possible is desirable to enhance the TE efficiency of the materials. The calculated $xx$ component of $\kappa_e$ is shown in Fig. \ref{fgr:zt-br}(c)–\ref{fgr:zt-i}(c), and the $yy$ component is given in Fig. S7-S8. Similar to $\sigma$, $\kappa_e$ is also dominated by electrons, which is undesirable. The $\kappa_e$ reaches a maximum of approximately $\SI{4.5}{}$ \klunit{}, due to electron carriers, in the Br-containing layers and $\SI{2.5}{}$ \klunit{} in the I-containing layers. The hole contributed $\kappa_e$ is notably lower than that due to electrons by a factor of 4 and 2.5 in Br-containing and I-containing layers, respectively. Such a high $\kappa_e$ due to electron carriers can significantly reduce the overall TE efficiency of SbXY JLs. However, the TE efficiency can be balanced by optimizing the contrasting quantity by improving the carrier concentrations in these layers. 

\subsection{\label{sec:TEeffi}Thermoelectric  efficiency}
The fundamental trade-off between $S$ and $\sigma$ defines the thermoelectric-derived quantity known as power factor $PF = S^2 \sigma$. A large $PF$ reflects the ability of a material to generate higher electrical power from a given temperature gradient $\Delta T$. Identifying the optimal carrier concentration, where the balance between $S$ and $\sigma$ yields the maximum $PF$, is therefore critical for designing efficient thermoelectric materials. The calculated $xx$ component of $PF$ is shown in Fig. \ref{fgr:zt-br}(d) and Fig. \ref{fgr:zt-i}(d) as a function of carrier concentration and temperature, while the $yy$ component is provided in Fig. S7-S8 in SM.

The $PF$ of SbXY layers is primarily governed by hole doping, with its maximum value attained at relatively low carrier concentrations. Although the $\sigma$ due to electron doping is dominant, the highest $PF$ arises from holes. Such a switch over from electron to hole doping stems from the fact that $S$ and $\sigma$ peak at different doping levels, preventing the realization of the maximum potential of $S$. To further clarify the hole dominance on $PF$ despite their lower mobility compared to electrons, the values of $S$ at the carrier concentrations where $ZT$ reaches its maximum are examined, cf. Fig. S9(c) in SM. The analysis shows that hole-derived $S$ is sufficiently larger than electron-derived $S$, thereby driving the maximum $PF$. A similar trend is observed for the $yy$ component of $PF$. Furthermore, the $xx$ component of $PF$ is roughly twice that of the $yy$ component, underscoring the anisotropic thermoelectric efficiency of SbXY layers. 

The peak values of the $xx$ component of $PF$ as a function of temperature are presented in Fig. S10 in SM. With increasing temperature, hole-derived $PF$ decreases, whereas electron-derived $PF$ increases, cf. Fig. S10 in SM. The combined influence of carrier concentration and temperature results in notable $PF$ values. In particular, maximum $PF$ values of \SI{3.0e-3}{} \pfunit{}, \SI{2.9e-3}{} \pfunit{}, \SI{1.4e-3}{} \pfunit{}, and \SI{2.5e-3}{} \pfunit{} are obtained for SbSeBr, SbTeBr, SbSeI, and SbTeI, respectively, under hole doping at \SI{300}{\kelvin}. Approximately half of these values are achieved through electron doping. The $PF$ values reported here are comparable to, though lower than, those of other 2D materials such as Pb$_2$SSe~\cite{Mamani2024}, Pb$_2$Sb$_2$S$_2$~\cite{YGan2021}, and (PbS)$_2$~\cite{PJin2022}. Notably, with the exception of SbSeI, electron contributions dominate the $PF$ while hole contributions are suppressed, thereby mitigating the bipolar effect. This selective suppression of bipolar transport enables SbXY layers to outperform many other 2D thermoelectrics~\cite{poonam2024jmca, CDing2023, Angran2025, MLIU2022}.

Finally, the thermoelectric efficiency, represented by the Figure of Merit $ZT$, is calculated as a function of doping concentration at different temperatures. The $ZT$ is defined as the normalized power factor $PF$ with respect to the total thermal conductivity $\kappa$ ($=\kappa_e + \kappa_l$) and scaled by the absolute temperature $T$, cf. Eq. \ref{eq:ZT}. The calculated $xx$ component of $ZT$ is shown in Fig. \ref{fgr:zt-br}(e) and Fig. \ref{fgr:zt-i}(e). In contrast to $PF$, $ZT$ is influenced by both electron and hole doping. Since experimental data for the melting temperature of SbXY JLs are not available, $ZT$ was evaluated up to \SI{1000}{\kelvin} as a representative case. Among SbXY layers, SbSeBr exhibits the best performance, yielding a modest $ZT$ of \SI{0.12}{} at 300 K and a maximum of 0.6 at 1000 K under hole doping. In SbSeI, electrons and holes contribute almost equally to thermal conversion, which may ultimately reduce the net efficiency due to the bipolar effect, tuning the SbSeI layer less favorable for practical applications. For the remaining layers, hole carriers dominate the thermal conversion and sustain a positive net yield. Moreover, consistent with the anisotropic $PF$, the $yy$ component of $ZT$ is significantly lower than the $xx$ component. These findings indicate that SbXY JLs could be useful in applications requiring directional suppression of thermal conduction, while simultaneously enabling efficient heat transport along another crystallographic axis. Overall, the results highlight SbXY JLs as promising candidates for high-temperature thermoelectric applications, achieving efficiencies exceeding 0.5.

\begin{figure}[tb]
    \centering
    \includegraphics[width=0.48\textwidth, clip, trim=0in 0in 0in 0in]{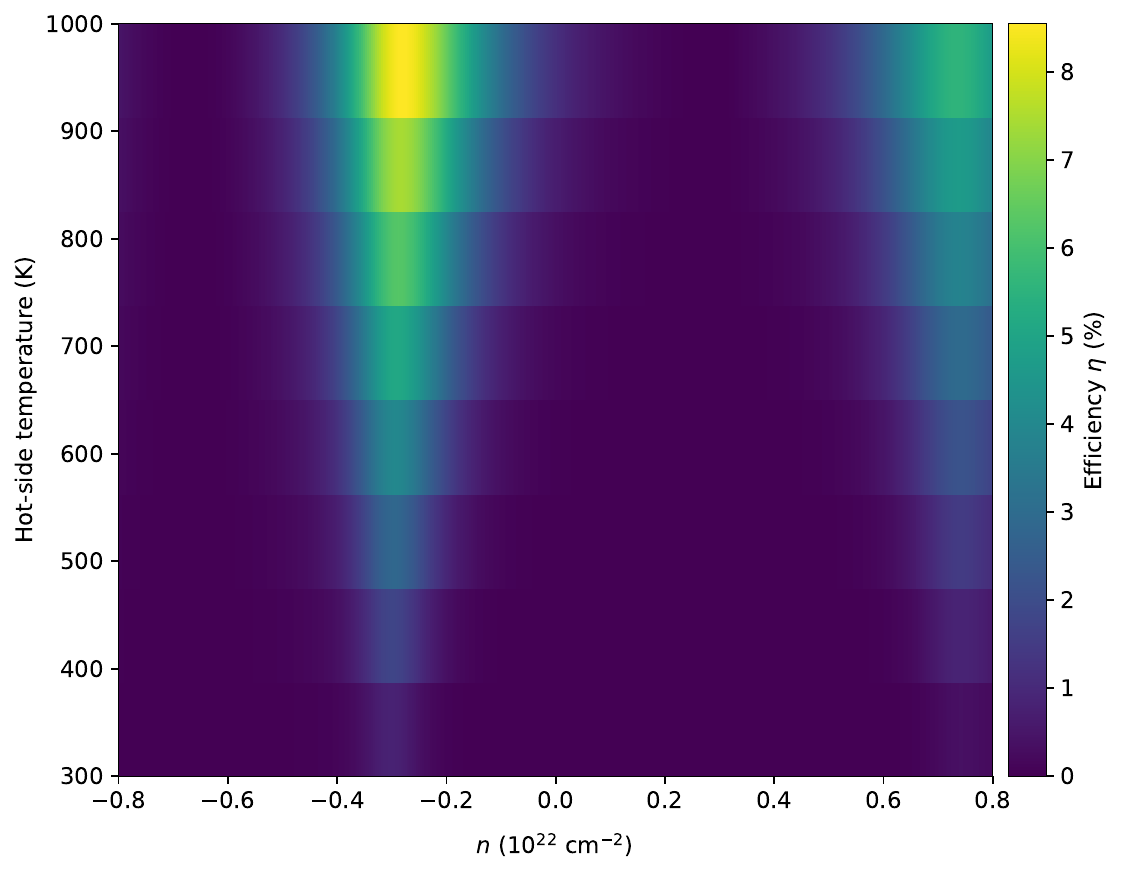}
    \includegraphics[width=0.48\textwidth, clip, trim=0in 0in 0in 0in]{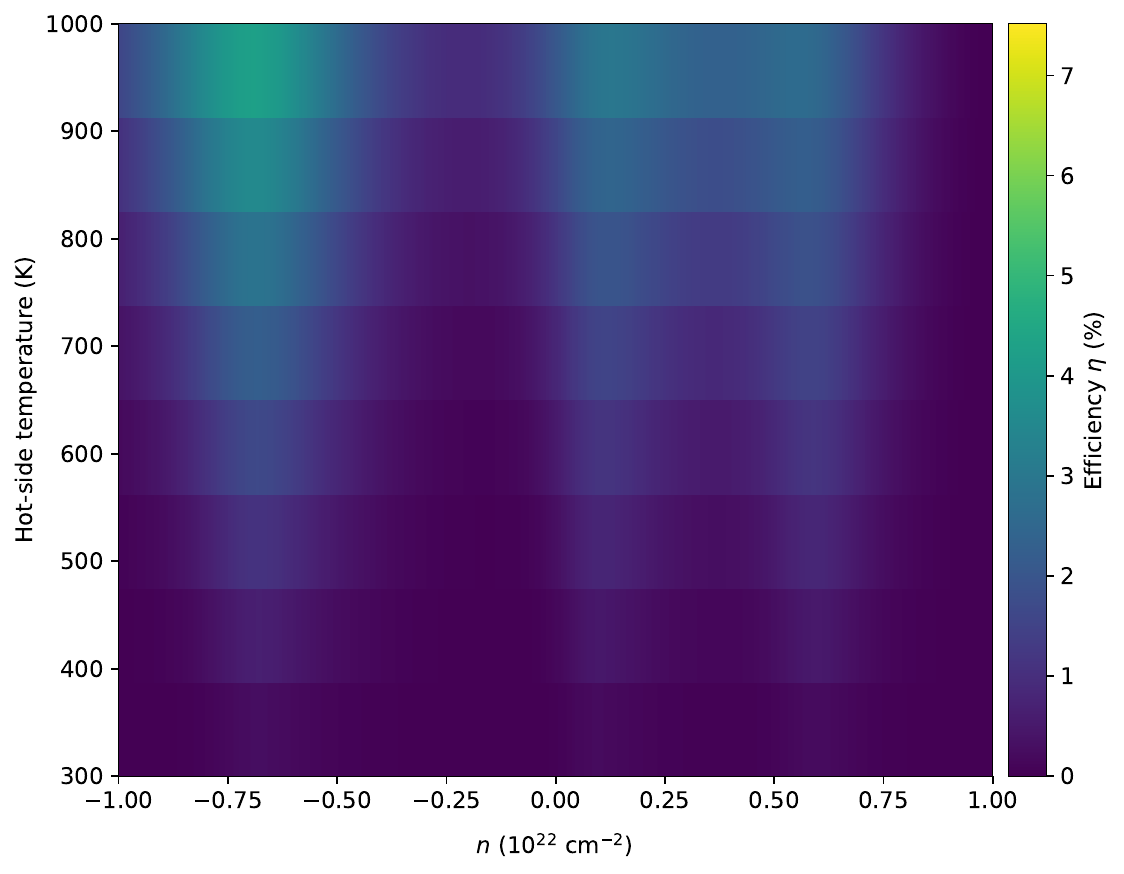}
    \caption{Calculated $xx$ component of Carnot efficiency $\eta$ for the conceptual TE device based on Br-containing layers SbSeBr and SbTeBr as function of hot side temperature under different doping concentration. The cold side temperature is fixed at 200 K. The SbTeBr layer results in better efficienty as better $\eta$ can be achieved at low hot side temperature.}
     \label{fig:eta-Br}
\end{figure}

\begin{figure}[tb]
    \centering
    \includegraphics[width=0.48\textwidth, clip, trim=0in 0in 0in 0in]{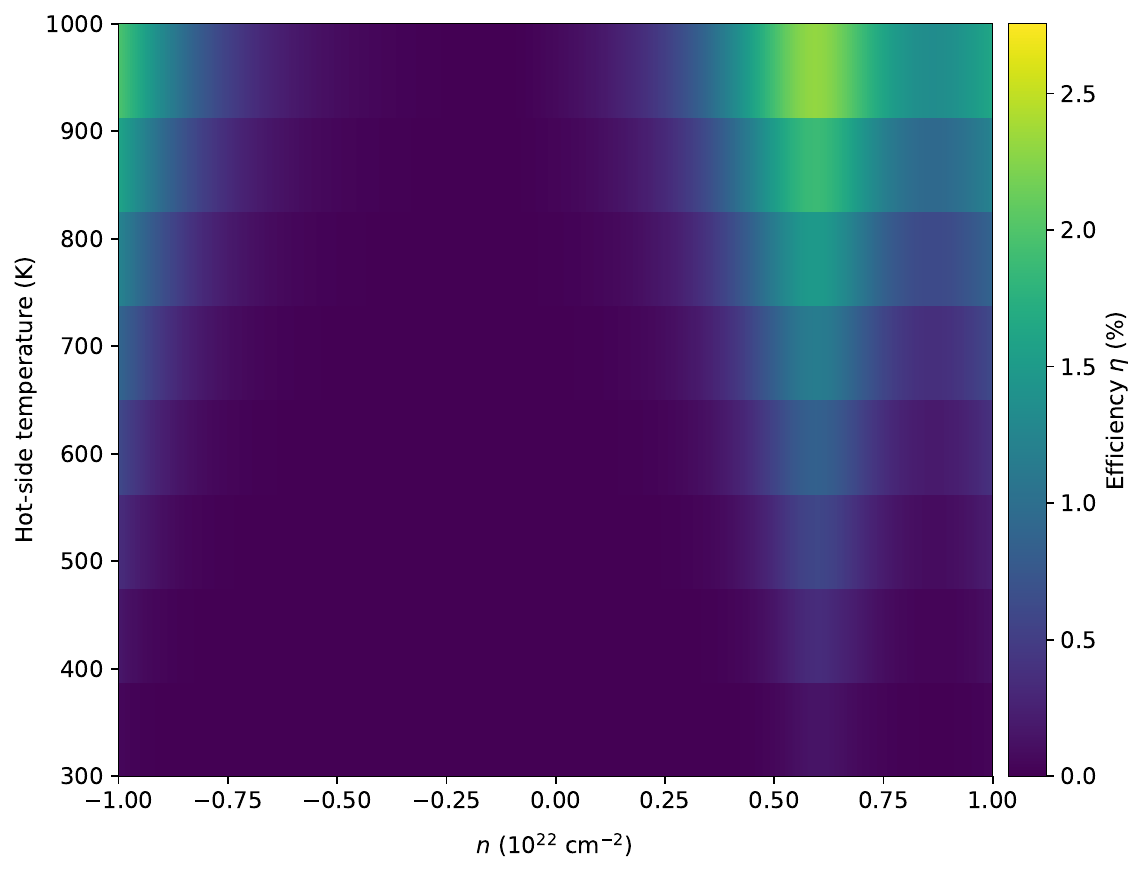}
    \includegraphics[width=0.48\textwidth, clip, trim=0in 0in 0in 0in]{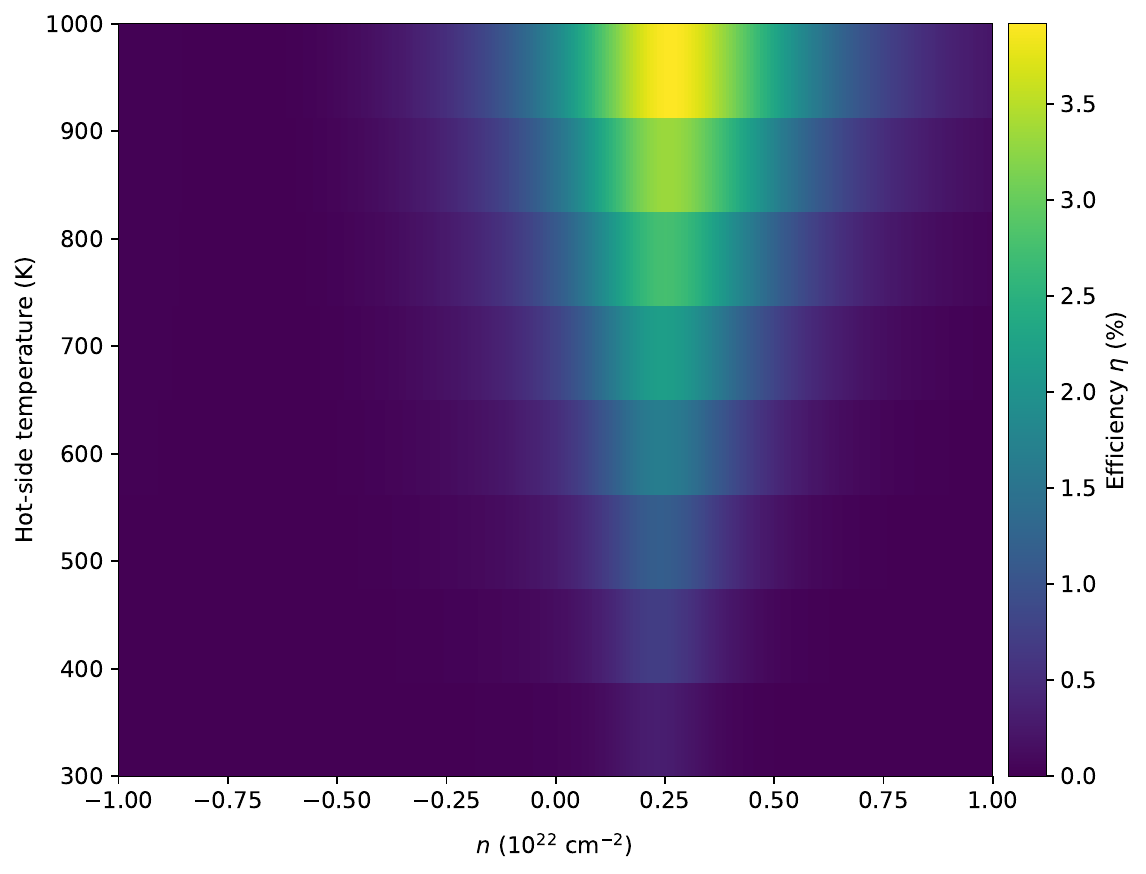}
    \caption{Calculated $xx$ component of Carnot efficiency $\eta$ for the conceptual TE device based on I-containing layers SbSeI and SbTeI as function of hot side temperature under different doping concentration. The cold side temperature is fixed at 200 K. The SbTeBr layer results in better efficienty as better $\eta$ can be achieved at low hot side temperature.}
     \label{fig:eta-I}
\end{figure}

\section{Conclusions}
In this study, spin-polarized first-principles calculations are employed to investigate the thermoelectric performance of SbXY (X = Se, Te; Y = Br, I) Janus layers. Ab initio molecular dynamics simulations confirm their structural stability, revealing that the 1T phase with space group symmetry \texttt{Pm31} remains stable up to 1000 K. This rules out the phase change possibility into other common layered phases, such as 2H or 3R (as observed in MoS$_2$), thereby establishing the robustness of SbXY JLs. The calculated mean-square displacement further demonstrates that atoms retain their equilibrium positions without significant structural distortions under thermal excitation. The thermal conductivity is evaluated using the force-displacement dataset from ab initio molecular dynamics. The pronounced ionic bonding of Sb–Br enhances long-range Coulomb interactions, facilitating strong coupling between optical and acoustic phonons via the Fr\"ohlich interaction. This coupling suppresses phonon lifetimes and significantly lowers the thermal conductivity of Br-containing layers compared to I-containing layers. Consequently, SbSeBr and SbTeBr exhibit reduced thermal conductivity relative to SbSeI and SbTeI, which is consistent with the determined shorter phonon lifetimes observed in the Br-based systems.

Apart from the structural stability analysis, the calculated electronic properties reveal that SbXY layers are indirect bandgap semiconductors, whose band gaps vary in the range of 1.1 eV – 1.3 eV, as confirmed by their electronic band structures. The density of states indicates that the valence band is primarily composed of $p$ orbitals from X and Y atoms, while the conduction band is dominated by Sb-$p$ orbitals. These orbital contributions result in nearly constant valence band maxima and conduction band minima when changing the X and Y combinations, which yield similar band gaps. Analysis of carrier effective masses underlines pronounced anisotropy in both carrier mobility and relaxation time. Electrons possess a low effective mass, thus being more mobile, indicating that transport is dominated by a single carrier type. The direction-dependent thermoelectric properties reveal superior performance along the $xx$ direction, whereas the $yy$ direction shows limited response due to reduced elastic behavior. While electrons dominate electrical conductivity due to their lower effective mass, the power factor $PF$ is governed by holes, as the Seebeck coefficient is higher for holes than electrons at the carrier concentrations which maximize $PF$. Though the holes are heavier, a competitive $PF$ on the order of $10^{-3}$ \pfunit{} is achieved, which is comparable to other 2D systems containing heavy elements such as Pb. These enhanced $PF$s assist in balancing the adverse effects of thermal conductivities due to phonons and charge carriers, which cause a moderate Figure of Merit $ZT$ at 300 K. With increasing temperature, $ZT$ increases steadily, reaching a maximum of 0.6 at 1000 K under hole doping in SbSeBr. In contrast, SbSeI suffers from bipolar conduction at elevated temperatures, reducing its efficiency and limiting its applicability in thermoelectric applications. Our findings recommend the SbXY JLs, particularly SbSeBr, as promising candidates for next-generation thermoelectric devices operating at elevated temperatures.

\section*{Conflicts of interest}
The authors declare no competing interests.

\begin{suppinfo}
The additional information such as the details of transport properties, mobility and relaxation time, effective mass, mean-square displacements, and thermoelectric properties along $yy$ direction are presented in the Supplementary Information
\end{suppinfo}

\begin{acknowledgement}
MV gratefully acknowledges Prof. Dirk C. Meyer for providing an excellent research environment at the Center for Efficient High-Temperature Processes and Materials Conversion (ZeHS), TU Bergakademie Freiberg. MV thanks the Department of Information Services and Computing at Helmholtz-Zentrum Dresden-Rossendorf for access to high-performance computing resources. Furthermore, MV and MZ also acknowledge computing time on the cluster of the Faculty of Mathematics and Computer Science at Technische Universit\"at Bergakademie Freiberg, operated by the University Computing Center (URZ) and funded by the Deutsche Forschungsgemeinschaft under grant number 397252409.
\end{acknowledgement}

\bibliography{reference}

\end{document}